\documentclass[aps,prb,twocolumn,floatfix,showpacs,numerical,footinbib,superscriptaddress]{revtex4-1}

\usepackage{graphicx}

\usepackage{amsfonts}
\usepackage{amsmath}
\usepackage{amssymb}

\usepackage[%
  colorlinks=true,
  urlcolor=blue,
  linkcolor=blue,
  citecolor=blue
]{hyperref}

\usepackage[pdftex]{color}

\newcommand{\sgn}[1]{\mathrm{sgn} \left( #1 \right)}
\newcommand{\e}{\mathrm{e}}
\newcommand{\im}{\mathrm{i}}
\newcommand{\di}{\mathrm{d}}
\newcommand{\ket}[1]{| #1 \rangle}
\newcommand{\bra}[1]{\langle #1 |}
\newcommand{\braket}[2]{\langle #1 | #2 \rangle}

\makeatletter
\DeclareFontFamily{OMX}{MnSymbolE}{}
\DeclareSymbolFont{MnLargeSymbols}{OMX}{MnSymbolE}{m}{n}
\SetSymbolFont{MnLargeSymbols}{bold}{OMX}{MnSymbolE}{b}{n}
\DeclareFontShape{OMX}{MnSymbolE}{m}{n}{
    <-6>  MnSymbolE5
   <6-7>  MnSymbolE6
   <7-8>  MnSymbolE7
   <8-9>  MnSymbolE8
   <9-10> MnSymbolE9
  <10-12> MnSymbolE10
  <12->   MnSymbolE12
}{}
\DeclareFontShape{OMX}{MnSymbolE}{b}{n}{
    <-6>  MnSymbolE-Bold5
   <6-7>  MnSymbolE-Bold6
   <7-8>  MnSymbolE-Bold7
   <8-9>  MnSymbolE-Bold8
   <9-10> MnSymbolE-Bold9
  <10-12> MnSymbolE-Bold10
  <12->   MnSymbolE-Bold12
}{}

\let\llangle\@undefined
\let\rrangle\@undefined
\DeclareMathDelimiter{\llangle}{\mathopen}%
                     {MnLargeSymbols}{'164}{MnLargeSymbols}{'164}
\DeclareMathDelimiter{\rrangle}{\mathclose}%
                     {MnLargeSymbols}{'171}{MnLargeSymbols}{'171}
\makeatother

\begin{document}

\title{Strongly angle-dependent magnetoresistance in Weyl semimetals with long-range disorder}

\author{Jan Behrends}
\affiliation{Max-Planck-Institut f\"ur Physik komplexer Systeme, 01187 Dresden, Germany}
\author{Jens H.\ Bardarson}
\affiliation{Max-Planck-Institut f\"ur Physik komplexer Systeme, 01187 Dresden, Germany}
\affiliation{Department of Physics, KTH Royal Institute of Technology, Stockholm, SE-106 91 Sweden}

\begin{abstract}
The chiral anomaly in Weyl semimetals states that the left- and right-handed Weyl fermions, constituting the low energy description, are not individually conserved, resulting, for example, in a negative magnetoresistance in such materials.
Recent experiments see strong indications of such an anomalous resistance response; however, with a response that at strong fields is more sharply peaked for parallel magnetic and electric fields than expected from simple theoretical considerations.
Here, we uncover a mechanism, arising from the interplay between the angle-dependent Landau-level structure and long-range scalar disorder, that has the same phenomenology.  
In particular, we analytically show, and numerically confirm, that the internode scattering time decreases exponentially with the angle between the magnetic field and the Weyl node separation in the large field limit, while it is insensitive to this angle at weak magnetic fields. 
Since, in the simplest approximation, the internode scattering time is proportional to the anomaly-related conductivity, this feature may be related to the experimental observations of a sharply peaked magnetoresistance.
\end{abstract}

\maketitle

Weyl semimetals are a recently realized family of materials\cite{Weng:2015ec,Huang:2015ig,Xu:2015kb,Lv:2015fj} in which the band dispersion is linear and crosses in points at the Fermi level\cite{Wang:2011hi}.
At low energy an effective description in terms of massless and chiral (left- or right-handed) relativistic Weyl fermions thus emerges.
The field theory of such fermions realizes the chiral anomaly\cite{Bell:1969dd,Adler:1969ir}, with left- and right-handed Weyl fermions not individually conserved.
In condensed-matter systems, where particle number conservation is manifest, this necessitates an equal number of fermions of each handedness;  Weyl fermions consequently always come in pairs\cite{Nielsen:1981ee,Nielsen:1981ki,Nielsen:1981kn}.
The anomaly is driven by non-orthogonal magnetic and electric fields that transmute one handedness into the other via bulk chiral Landau levels\cite{Nielsen:1983ce}.
Weyl semimetals are also topological, which leads to stable surface states with Fermi arcs\cite{Turner:2013cj,Hosur:2013ef}.
The combination of relativistic Weyl dispersion, chiral anomaly and Fermi arcs results in various distinctive transport properties, such as quantum oscillations from Fermi arcs\cite{Potter:2014cg,Zhang:2016dx,Bulmash:2016bw}, resonant transparency\cite{Baum:2015jl}, and nonlocal valley currents\cite{Parameswaran:2014cw}.

A further experimentally relevant consequence of the chiral anomaly, a large negative magnetoresistance, was realized by Nielsen and Ninomiya and discussed in their lucid 1983 paper\cite{Nielsen:1983ce}.
The longitudinal magnetoconductivity takes the form
\begin{equation}
 \sigma_{ii} = f (B)\,\tau_v
 \label{eq:conductivity}
\end{equation}
with $f(B) \propto B$ in the quantum limit with quantized Landau levels\cite{Nielsen:1983ce} and $f(B) \propto B^2$ in the semiclassical limit\cite{Son:2013kd,Morimoto:2016kw}, where this quantization can be neglected.
Importantly, it is the internode scattering time $\tau_v$---the time for scattering between different chiralities, which is generally much larger than the intranode chirality preserving scattering time $\tau_c$---that enters, resulting in an unusually large magnetoresistance.
Early transport experiments have obtained results consistent with this prediction\cite{Xiong:2015kl,Shekhar:2015io,Li:2015bz,Xiong:2016bx,Li:2016bj,Li:2016hs,Arnold:2016hp} (albeit in some cases interpreted as a result of current jetting\cite{Arnold:2016hp}), although not all features of these experiments are fully understood.
Measurements for Na$_3$Bi, where two Dirac nodes split up into four Weyl nodes due to Zeeman coupling, show a strong angular dependence of the conductivity\cite{Xiong:2015kl}, which at weak magnetic field is consistent with expectations, but at large magnetic field it is even much more strongly peaked at parallel electric and magnetic fields than can be explained assuming an angle-independent internode scattering time\cite{Burkov:2016kx}.

In this work we address this observation by studying the angular dependence of the magnetoresistance in the presence of long-range disorder, extending previous work on the interplay of long-range disorder and magnetic field in Weyl semimetals\cite{Pesin:2015kp,Lu:2017gq,Zhang:2016hu,Chen:2016gq}.
We show analytically that, within the Born approximation, the internode scattering time $\tau_v$ is exponentially reduced when the magnetic field is tilted away from the momentum-space separation of the Weyl nodes\footnote{The angle defined here differs from the experiments, where the magnetic and electric fields are tilted away from each other.}.
We limit ourselves to the simplest case of two Weyl fermions of opposite chirality separated in momentum space by the vector~$\mathbf{b}$.
The essential reason for the observed effect is understood from noting that tilting the magnetic field $\mathbf{B} = B \,\mathbf{e}_r$ away from $\mathbf{b}$ reduces the effective node separation to $b_r = \mathbf{b} \cdot \mathbf{e}_r$.
In the presence of long-range disorder, the internode scattering time decays rapidly with decreasing node separation, since this corresponds to a large momentum transfer, resulting, via the relation~\eqref{eq:conductivity}, to sharply peaked magnetoresistance.
The long-range nature of the disorder potential is essential---for short-range disorder the separation in momentum space does not influence the scattering rate\cite{Parameswaran:2014cw}.
In addition to this effect, there is a slightly more subtle effect due to the shift of the Landau-level wave-function-center with momentum, which affects the internode scattering time in the opposite direction by increasing the real space distance between the closest states in momentum space.
Remarkably, these two effects counterbalance each other at low magnetic field (magnetic length $\ell_B \gg \xi$ the disorder correlation length) such that the internode scattering time is independent of the angle. 
At large magnetic field ($\ell_B \ll \xi$) the shift of the node separation dominates and results in the aforementioned exponential decrease of the internode scattering time $\tau_v$. 

The argument just given is valid for low energies where the chemical potential $\mu < \sqrt{2}\,\hbar v/\ell_B$ such that one can ignore all but the lowest Landau level.
In this limit we provide explicit analytical results, and further extend them to higher energies, with the results remaining qualitatively the same, by taking into account scattering between different Landau levels.
This provides the detailed chemical potential dependence of the magnetoresistance.
Finally, these quantum limit results are confirmed by a numerical computations of the conductance at zero energy, using a transfer matrix technique\cite{Bardarson:2007iu}.

Given that we are interested in the internode scattering time $\tau_v = -\hbar / \mathrm{Im} \Sigma^R_{\chi\neq \chi'} $, with $\chi$ denoting the chirality, we need to compute the self-energy $\Sigma$.
The Weyl Hamiltonian describing two Weyl nodes in a magnetic field is 
\begin{equation}
 \mathcal{H} = v \left( \hbar \mathbf{k} + e\,\mathbf{A} \right) \cdot \boldsymbol\sigma\,\tau_z + V,
 \label{eq:weyl_hamiltonian}
\end{equation}
where $\boldsymbol\sigma$ is the vector of Pauli matrices and $\hbar \mathbf{k}$ is the momentum measured from the Weyl points that are separated by a vector $\mathbf{b} = b\,\mathbf{e}_z$.
The Pauli matrices $\tau_\mu$ act in the space of chiralities.
The disorder potential $V$ has Gaussian correlations
\begin{equation}
 \left\llangle  V (\mathbf{r} )\,V( \mathbf{r}') \right\rrangle = \frac{K_0}{(2\,\pi)^{3/2}} \,\left( \frac{\hbar v}{\xi} \right)^2 \exp \left[ - \frac{\left| \mathbf{r} - \mathbf{r}' \right|^2}{2\,\xi^2} \right]
 \label{eq:disorder_correlation}
\end{equation}
of dimensionless strength $K_0$, with $\llangle \cdot \rrangle$ denoting the disorder average.
The vector potential $\mathbf{A}$ lies in the $y$--$z$-plane with
\begin{align}
 \mathbf{B} &= B\,\mathbf{e}_r,
 &\mathbf{A} &= B\,x\,\mathbf{e}_\theta \\
 \mathbf{e}_r &= \cos \theta \,\mathbf{e}_z + \sin \theta \,\mathbf{e}_y,
 &\mathbf{e}_\theta & = \cos \theta\,\mathbf{e}_y - \sin \theta \,\mathbf{e}_z .
\end{align}
In a clean system $V=0$, the two chiralities are decoupled and their eigenfunctions can be found separately.
Rotating the $2\times 2$ Hamiltonian of chirality $\chi$ allows us to express it in terms of creation and annihilation operators in a Landau-level basis\cite{Ashby:2013kh,Meng:2016di}
\begin{equation}
 \mathcal{H}_\chi = \chi \frac{\hbar v}{\ell_B}\,\left( \begin{array}{cc} \ell_B\,k_r & \im\,\sqrt{2}\,a_{k_\theta} \\ -\im\,\sqrt{2}\,a_{k_\theta}^\dagger & - \ell_B\,k_r
 \end{array} \right),
 \label{eq:ll_hamiltonian}
\end{equation}
with the magnetic length $\ell_B = \sqrt{\hbar / (e B)}$, rotated momenta $k_{r,\theta} = \mathbf{k} \cdot \mathbf{e}_{r,\theta}$, and annihilation operator
\begin{equation}
  a_{k_\theta} = \frac{1}{\sqrt{2}} \left( \frac{x}{\ell_B} + \ell_B\,k_\theta + \im\,\ell_B\,k_x \right).
\end{equation}
The energy spectrum consists of dispersive Landau levels, which at positive energy are given by
\begin{equation}
 E_{n>0}^\chi =  \hbar \omega_B\,\sqrt{2\,n + \ell_B^2 k_r^2},~ E_0^\chi = \chi\,\hbar v\, k_r
 \label{eq:energy_levels}
\end{equation}
where the cyclotron frequency $\omega_B = v/\ell_B$. The corresponding eigenstates are\cite{Klier:2015bu,Abrikosov:1998be}
\begin{align}
 \ket{\Phi_{n> 0\,\mathbf{k}_\parallel}^\chi} &=\frac{1}{\sqrt{N_\chi}} \left(\frac{-\im\,\sqrt{2\,n}\, \ket{n-1} }{\ell_B\,k_r - \chi \sqrt{2\,n +\ell_B^2 k_r^2}},  \ket{n}\right)^T 
 \label{eq:eigenstates}
 \\
 \ket{\Phi_{0\,\mathbf{k}_\parallel}^\chi} &= \left(0, \ket{0} \right)^T
 \label{eq:eigenstates0}
\end{align}
with $N_\chi= 2\,\sqrt{2\,n+ \ell_B^2 k_r^2}/({\sqrt{2\,n + \ell_B^2 k_r^2}  - \chi \ell_B k_r})$ a normalization constant.
The wave functions describing the eigenstates of the number operator $a_{k_\theta}^\dagger a_{k_\theta} \ket{n} = n\,\ket{n}$,  $\braket{x,\mathbf{k}_\parallel}{n} = \psi_n \left( x /\ell_B + \ell_B\,k_\theta \right)/\sqrt{\ell_B}$, are proportional to the Hermite functions $\psi_n$ centered at $x = -\ell_B^2\,k_\theta$, such that the position along $x$ and the momentum along $\mathbf{e}_\theta$ are locked.
While the energies depend only on the single momentum component $k_r$, the wave functions depend only on $k_\theta$.
Lines of equal energy are parallel to $k_y$ when $\theta=0$ and tilt towards $k_z$ when $\theta$ increases\cite{Bulmash:2016bw}.

\begin{figure}
 \includegraphics[width=\linewidth]{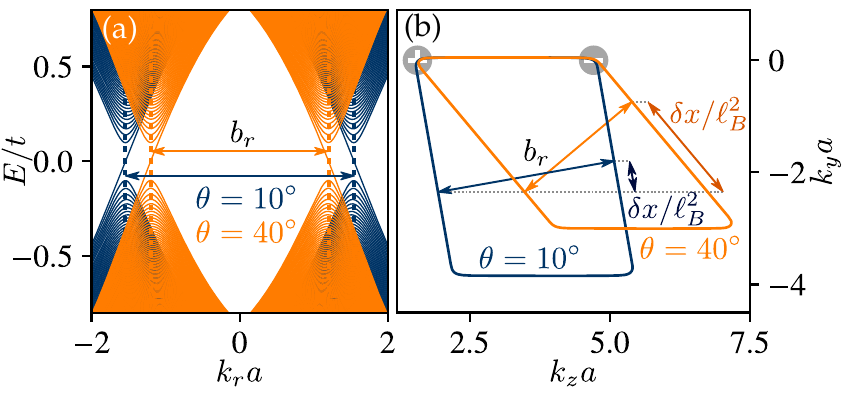}
 \caption{(a) Energy dispersion of a tight binding system\cite{Yang:2011im}, specified in Appendix~\ref{sec:tight_binding}, with two Weyl nodes separated by $\mathbf{b} = \pi\,\mathbf{e}_z$, as a function of $k_r$, and (b) Fermi surface at a chemical potential $\mu=-0.05\,t$ and a magnetic field of $\mathbf{B} = B\,\cos \theta\,\mathbf{e}_z + B\,\sin \theta \,\mathbf{e}_y$ for two different angles $\theta$ with $\ell_B = 14.1\,a$, and $a$ the lattice constant.
 The system is finite in the $x$ direction with $L=768$ sites and infinite in the other two directions.
 The dashed lines in the energy dispersion (a) show the position of the zeroth Landau level at zero energy expected from the continuum model, at $k_r a= \pm \pi/2\,\cos \theta$.
 The positions of the Weyl nodes in absence of a magnetic field are indicated by gray circles in (b).
 }
 \label{fig:different_theta}
\end{figure}

We corroborate these low energy results by calculating the band structure in a tight-binding system with a low-energy bulk dispersion that is well reproduced by two isolated Weyl nodes\cite{Yang:2011im}.
The Fermi-arc structure\cite{Bulmash:2016bw} is captured by taking the system finite in the $x$ direction.
A magnetic field in the $y$-$z$ plane changes the linear dispersion to a Landau-level structure with momentum-space separation of the chiralities $b_r = b\,\cos \theta$ [see Fig.~\ref{fig:different_theta}(a)].
At low energies $\mu < \sqrt{2} \hbar \omega_B$, the Fermi surface consists of the two zeroth Landau levels (tilted lines) that are connected by Fermi arcs (at fixed $k_y$) [see Fig.~\ref{fig:different_theta}(b)].
As the zeroth Landau levels tilt towards $k_z$ with increasing angle between the magnetic field and node separation, the distance in real space $\delta x = \ell_B^2\,\delta k_\theta$, with $k_\theta$ measured from the Weyl nodes in absence of a magnetic field, simultaneously increases for those states with minimal momentum-space distance.

The essential reason for the enhanced internode scattering can now be intuitively understood from the band structure:
for a disorder potential with long-range correlations, the amplitude for internode scattering increases when the magnetic field is tilted away from the Weyl-node momentum axis, due to the reduced momentum-space distance $b_r$, while simultaneously decreasing because of the increased real-space distance $\delta x$.
To substantiate this argument, we explicitly calculate the self-energy.

\begin{figure}
 \includegraphics[width=.95\linewidth]{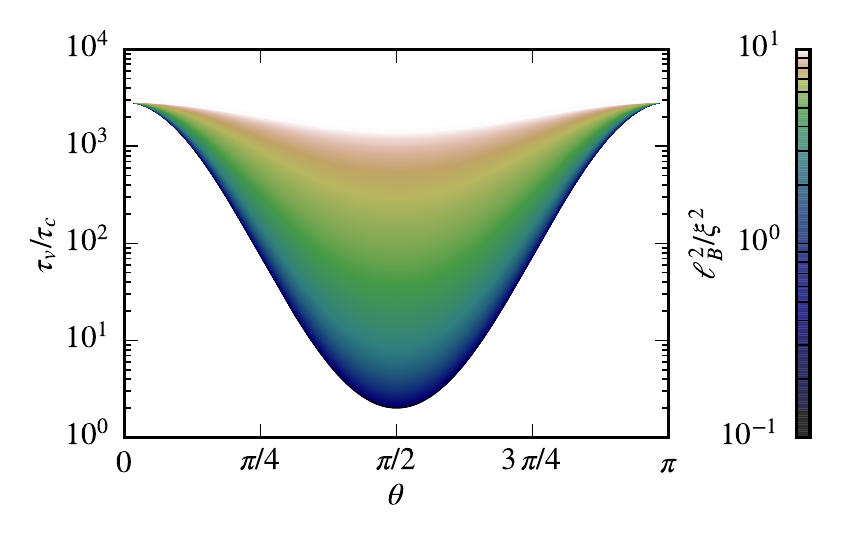}
 \caption{Ratio of inter- and intranode scattering times $\tau_v/\tau_c$, for two isolated Weyl nodes in the presence of long-range disorder, as a function of the angle  $\theta$ between the node separation $\mathbf{b}$ and the externally applied magnetic field $\mathbf{B}$.
 The ratio is evaluated at $\mu=0$ with a node separation $\xi \,b = 4$.
 The color scale shows different magnetic field strengths given by $\ell_B^2/\xi^2$, with the two limits $\ell_B/\xi \to 0$ and $\ell_B/\xi \to \infty$ described in the main text.
 }
 \label{fig:scattering_time}
\end{figure}

The disorder-induced self-energy correction for a Landau level of certain chirality can be split up into two contributions:
scattering within the chirality, which defines the intranode scattering time $\tau_c$, and scattering between different chiralities, which defines the internode scattering time $\tau_v$.
These corrections are expressed to lowest order in the disorder potential via the Born approximation, with $\mathbf{k}_\parallel = (k_r,k_\theta)$,
\begin{equation}
 \Sigma^{m}_{\chi\chi'}  (\im\,\omega_n,\mathbf{k}_\parallel) = \sum_{m',k_r'} \mathcal{G}_{m' \chi'} (\im\,\omega_n , k_r') \sum_{k_\theta'} \Gamma_{mm'}^{\chi\chi'}
 \label{eq:self_energy}
\end{equation}
with the disorder correlation in the Landau-level basis
\begin{equation}
 \Gamma_{mm'}^{\chi\chi'} = \left\llangle \bra{\Phi_{m\,\mathbf{k}_\parallel}^{\chi} } V \ket{\Phi_{m'\,\mathbf{k}_\parallel'}^{\chi'} } \bra{\Phi_{m'\,\mathbf{k}_\parallel'}^{\chi'} } V \ket{\Phi_{m\,\mathbf{k}_\parallel}^{\chi} } \right\rrangle
\end{equation}
and the Green's function of the clean system $\mathcal{G}_{m \chi} (\im\,\omega_n , k_r) = 1/(\im\,\omega_n - E_m^\chi)$.
The splitting up of the momentum summation in this expression is possible since the energies only depend on the momentum component $k_r'$.
The diagonal elements of $\Sigma$ with $\chi = \chi'$ give intranode scattering, while off-diagonal elements with $\chi \neq \chi'$ give internode scattering.
Inserting the Gaussian correlation of Eq.~\eqref{eq:disorder_correlation},
\begin{align}
 \Gamma_{m m'}^{\chi\chi'}	=& \frac{K_0\,\hbar^2 v^2}{\sqrt{2\,\pi}\,L^2} \e^{-\frac{1}{2} \,\left(\xi\, \delta \mathbf{q}_\parallel^{\chi\chi'} \right)^2}\int \di x\,\di x'\,\e^{-\frac{(x-x')^2}{2\,\xi^2}}  \label{eq:gamma_mm} \\
				& \times \Phi^{\chi\,\dagger}_{m\,\mathbf{k}_\parallel} (x) \,\Phi^{\chi'}_{m'\,\mathbf{k}_\parallel'} (x)\,\Phi^{\chi'\,\dagger}_{m'\,\mathbf{k}_\parallel'} (x') \,\Phi^{\chi}_{m\,\mathbf{k}_\parallel} (x'), \nonumber
\end{align}
where $\delta\mathbf{q}_\parallel^{\chi\chi'} = \mathbf{k}_\parallel - \mathbf{k}_\parallel' - (\chi - \chi') \mathbf{b}/2$ is the momentum transfer between the involved states.
Scattering is therefore dominated by small momentum transfer $\delta \mathbf{q}_\parallel^{\chi\chi'} \approx 0$.
Note that $\Phi^{\chi}_{n\,\mathbf{k}_\parallel} (x) = \braket{x}{\Phi^\chi_{n\,\mathbf{k}_\parallel}}$ are two-component wave functions.
In the ultra-quantum limit at energies $\mu < \sqrt{2}\,\hbar \omega_B$, it is sufficient to compute $\Gamma_{m m'}^{\chi\chi'}$ by including scattering only between zeroth Landau levels.
With the wave functions~\eqref{eq:eigenstates0}, the disorder correlations~\eqref{eq:gamma_mm} simplifies to
\begin{equation}
 \Gamma_{0 0}^{\chi\chi'}
 = \frac{K_0\,\hbar^2 v^2}{\sqrt{2\,\pi}\,L^2}\, \frac{\xi}{\sqrt{\ell_B^2 + \xi^2}} \e^{-\frac{1}{2} \left(  \xi\, \delta\mathbf{q}_\parallel^{\chi\chi'} \right)^2 -\frac{1}{2}\,\ell_B^2 \,\left(k_\theta -k_\theta' \right)^2} .
\end{equation}
The self-energy correction for scattering within the same chirality is now straightforward.
Integration over momenta $\mathbf{k}_\parallel'$ results in an inverse intranode scattering time
\begin{equation}
  \frac{1}{\tau_c}	= -\frac{1}{\hbar} \mathrm{Im} \left. \Sigma^{0\,R}_{\chi\chi} \right|_{k_r = \chi \frac{\varepsilon}{\hbar v}}
			= \frac{K_0}{4\,\pi}\,\frac{v}{\xi} \,\frac{\xi^2}{\ell_B^2+ \xi^2}.
\end{equation}
For the internode scattering time the momentum difference between the chiralities is enlarged by the node separation, resulting in
\begin{equation}
 \frac{1}{\tau_v} = \frac{K_0}{4\,\pi}\,\frac{v}{\xi} \,\frac{\xi^2}{\ell_B^2+ \xi^2}\,\e^{-\frac{1}{2} \xi^2 \,\left[ \left(\frac{2\,\mu}{\hbar v} - b_r\right)^2 - \frac{\ell_B^2}{\ell_B^2 + \xi^2}\,b_\theta^2 \right]}
 \label{eq:tau_v}
\end{equation}
with $b_{r,\theta} = \mathbf{b} \cdot \mathbf{e}_{r,\theta}$.
The ratio of inter- and intranode scattering time at zero energy is therefore
\begin{equation}
 \frac{\tau_v}{\tau_c} = \exp \left[ \frac{1}{2} \xi^2 \, b^2 \,\left( \cos^2 \theta + \frac{\ell_B^2}{\ell_B^2 + \xi^2}\,\sin^2 \theta \right) \right] .
 \label{eq:scattering_ratio}
\end{equation}
In the limit of small magnetic fields, $\ell_B \gg \xi$, the ratio does not depend on the angle $\theta$ between the node separation and magnetic field.
In contrast, for large magnetic fields, $\ell_B \ll \xi$, the ratio decreases exponentially when the magnetic field is tilted away from the node separation.
This functional behavior is illustrated in Fig.~\ref{fig:scattering_time}, and constitutes one of our main results.

To extend our results to energies larger than $\sqrt{2}\,\hbar \omega_B$, where higher Landau levels are occupied, a full solution for the disorder correlation~\eqref{eq:gamma_mm} is needed.
The integrals over both momentum components are similar to those appearing in the treatment of the integer quantum Hall effect\cite{Xie:1990gv}, resulting in analogous expressions.
Since the full analytical expression is too complicated to give additional insight, we relegate its display to Appendix~\ref{sec:appendix_all_states}.

\begin{figure}
 \includegraphics[width=\linewidth]{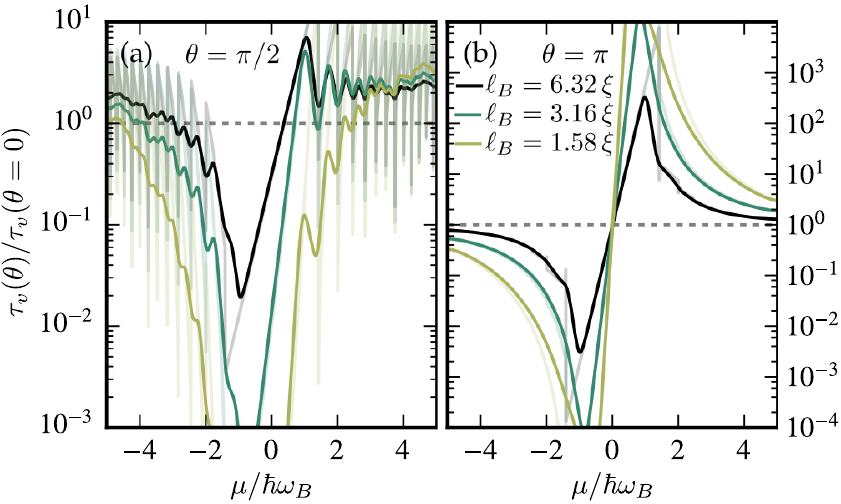}
 \caption{Ratio of internode scattering times $\tau_v (\theta)/\tau_v (\theta=0)$ at different tilt angles $\theta$ as a function of the chemical potential $\mu$, evaluated away from the ultra quantum limit, at (a) $\theta = \pi/2$ and (b) $\theta = \pi$.
 The opaque lines show temperature-broadened results (with $\tau_v (\mu) = -\int \di \omega \tau_v^{T=0} (\omega) \partial n_F /\partial \omega$ and $T = 0.067\,\hbar \omega_B$) for a better visibility of the overall trend, while the semitransparent lines in the background show zero-temperature results.
 The node separation at zero field is fixed to $b\,\xi = 10$.
 The dashed gray line depicts a ratio of $1$ as a guide for the eyes.
 }
 \label{fig:scattering_time_ratio}
\end{figure}

In Fig.~\ref{fig:scattering_time_ratio}(a) we show the ratio of the internode scattering times with the node separation orthogonal and parallel to the magnetic field, $\tau_v (\theta = \pi/2) / \tau_v (\theta=0)$, as a function of the chemical potential for different magnetic field strengths.
As in the ultra quantum limit, the change of the internode scattering time is more pronounced for large fields.
For small $\ell_B/\xi$, the decrease persists up to higher energies than for large $\ell_B/\xi$.
The behavior for energies in the ultra-quantum limit with $\mu < \sqrt{2} \hbar \omega_B$ can be best understood from Eq.~\eqref{eq:tau_v}, which gives a linearly increasing logarithm of the ratio due to the reduced momentum-space distance with increasing energy.
In Fig.~\ref{fig:scattering_time_ratio}(b), we further compare the ratio of scattering times for the node separation anti-parallel and parallel to the applied field.
These two scattering times are not identical due to the dispersion of the zeroth Landau levels:
their momentum-space separation is reduced when $\mu > 0$, while changing the direction of the magnetic field $\mathbf{B} \to - \mathbf{B}$ reverses this effect and increases the node separation when $\mu>0$.
This effect could be visible in experiments since other reasons for an angle-dependent magnetoresistance, such as non-isotropic Fermi velocities, do not change upon reversing the magnetic field direction.

\begin{figure}
 \includegraphics[width=\linewidth]{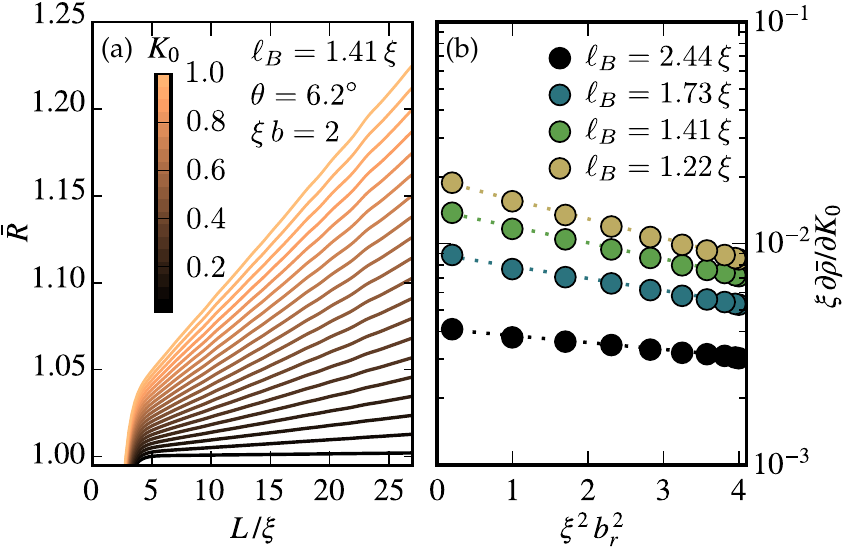}
 \caption{Numerically obtained resistance for two isolated Weyl cones with long-range internode disorder at zero chemical potential.
 (a) The disorder-averaged dimensionless resistance $\bar{R}$ normalized by the number of modes increases with the system size with constant slope $\bar{\rho} = \partial \bar{R}/\partial L$.
 The slope is proportional to the inverse scattering time $\tau_v^{-1}$ and therefore linearly increases with disorder strength $K_0$.
 (b) The logarithm of the derivative $\xi\,\partial \bar{\rho}/\partial K_0$ scales quadratically with $\xi\,b_r$.
 Circles represent numerical results, and the dotted line shows the results from the analytical expectation given in Eq.~(\ref{eq:analytical_expectation}).
 All numerical data are averaged over 200 different disorder realizations.
 }
 \label{fig:transfer_results}
\end{figure}

We verify the approximation that the internode scattering time is the relevant transport time, and thereby our results, by numerically calculating the conductance using a transfer matrix representation\cite{Bardarson:2007iu,Xypakis:2017ba} of the solutions of the Weyl equation $\mathcal{H}\,\psi = E\,\psi$, previously employed for disordered Weyl nodes in the absence of a magnetic field\cite{Sbierski:2014bo,Sbierski:2015ii,Sbierski:2017ix,Trescher:2017ka}.
A random scattering potential with correlations as in Eq.~(\ref{eq:disorder_correlation}) is transformed into the Landau-level basis and added to the Hamiltonian in terms of the raising and lowering operators, Eq.~(\ref{eq:ll_hamiltonian}).
The details of this approach are given in Appendix~\ref{sec:transfer_matrix}.
With this method, we obtain the matrix of transmission amplitudes $t$ for a system that has periodic boundary conditions in the transverse $x$ and $y$ directions and a finite length $L$ in the $z$ direction; the conductance is then given by the Landauer formula
\begin{equation}
 G = \frac{e^2}{h} \mathrm{Tr} \left[ t^\dagger t \right] .
\end{equation}
In Fig.~\ref{fig:transfer_results}(a), we show the dimensionless resistance $\bar{R} = G^{-1}\,N\,e^2/h$, normalized by $N$ transverse modes, for fixed node separation $b\xi$ and ratio $\ell_B/\xi$, as a function of system length for various disorder strengths.
The resistance increases linearly with $L$, giving a constant resistivity $\bar{\rho} = \partial \bar{R} /\partial L$.
For small disorder strength $K_0 \lesssim 1$, $\bar{\rho}$ increases linearly with $K_0$.
Figure~\ref{fig:transfer_results}(b) shows the dimensionless quantity $\xi\,\partial \bar{\rho}/\partial K_0$ as a function of the squared effective node separation $b_r^2 = b^2 \cos^2 \theta$, where we included the analytical expectation 
\begin{equation}
 \xi \frac{\partial \bar{\rho}}{\partial K_0} = 2\,\frac{1}{K_0}\,\frac{\xi}{v}\, \tau_v^{-1}
 \label{eq:analytical_expectation}
\end{equation}
as a guide for the eye.
The good agreement provides a numerical confirmation of the main finding of this work:
magnetotransport is not just dominated by the size of the node separation itself, but also by the tunable angle between applied magnetic field and the node separation.

In conclusion, we find that the internode scattering time in Weyl semimetals at zero energy exponentially decreases when an external magnetic field is tilted away from the Weyl node separation.
For small magnetic fields, this drop vanishes and the internode scattering time is angle independent.
Numerically, we confirm that the internode scattering time is the relevant transport time and proportional to the conductivity.
Away from the ultra-quantum limit, where transport is dominated by the zeroth Landau level, the internode scattering time remains angle dependent, and decreases for large magnetic fields tilted away from the Weyl node separation.
The angle-dependent internode scattering time may be related to the experimentally observed sharply peaked magnetoresistance at large magnetic fields.
Our predictions can, in principle, be experimentally confirmed by measuring the conductivity along the axis of parallel electric and magnetic fields with all fields tilted against the node separation.
The predictions can be extended for a larger number of Weyl nodes. The behavior presented in this work especially applies when the separation between pairs of Weyl nodes is smaller than the separation of the pairs, cf.\ Appendix~\ref{sec:experimental_signatues}.
The difference in conductivity for magnetic fields parallel and anti parallel to the node separation is another experimentally accessible signature.

\acknowledgments
We thank Adolfo G. Grushin and Bj\"{o}rn Sbierski for useful discussions. This work was supported by the ERC Starting Grant No.~679722.

\appendix
\section{Tight-binding Hamiltonian}
\label{sec:tight_binding}

The Hamiltonian used for the tight-binding results in the main text is\cite{Yang:2011im}
\begin{equation}
 \mathcal{H} = t \left[ \sin ( k_x a) \,\sigma_x + \sin (k_y a) \,\sigma_y \right] + M_\mathbf{k}\,\sigma_z
\end{equation}
with $M_\mathbf{k} = t\,\left[ 2- \cos (k_x a) - \cos (k_y a)\right] + t \,\cos (k_z a)$, realized on a cubic lattice with lattice constant $a$.
Its energy dispersion exhibits two isotropic Weyl nodes with velocity $t$ at $k_z a = \pm \pi/2$.
A magnetic field in the $y$-$x$-plane is introduced via minimal coupling.

\section{Scattering between different Landau levels}
\label{sec:appendix_all_states}

To include scattering between Landau levels $m$ and $m'$ in the self energy calculation, we insert the eigenspinors
\begin{align}
 \ket{\Phi_{n> 0\,\mathbf{k}_\parallel}^\chi} &=\frac{1}{\sqrt{N_\chi}} \left(\frac{-\im\,\sqrt{2\,n}\, \ket{n-1} }{\ell_B\,k_r - \chi \sqrt{2\,n +\ell_B^2 k_r^2}},  \ket{n}\right)^T 
 \\
 \ket{\Phi_{0\,\mathbf{k}_\parallel}^\chi} &= \left(0, \ket{0} \right)^T
\end{align}
into the expression for the disorder correlator in Landau level basis to obtain 
\begin{align}
 \sum_{k_\theta'} \Gamma_{m\,m'}^{\chi\chi'}
	&= \frac{K_0}{2\,\pi\,L} \frac{\hbar^2 v^2}{\xi}\,\frac{\xi^2}{\ell_B^2+\xi^2} \frac{\e^{-\frac{1}{2} \xi^2 \delta k_r^2}}{N_m\,N_{m'}} \label{eq:gamma_mm_full} \\
	& \times  \left[ I_{m,m'}^{m,m'} + 2\,\zeta\, I_{m,m'}^{m-1,m'-1} + \zeta^2\,I_{m-1,m'-1}^{m-1,m'-1} \right], \nonumber
\end{align}
where the integrals
\begin{align}
 I_{nm}^{n'm'}
 =	& \frac{\xi}{\sqrt{2 \pi}} \,\frac{\ell_B^2+\xi^2}{\xi^2} \int \di k_\theta' \,\di x\,\di x'\,\e^{-\frac{1}{2} \xi^2 \delta k_\theta^2 -  \frac{(x-x')^2}{2\,\xi^2}} \nonumber \\
	& \times \phi_{n\,k_\theta} (x) \phi_{m\,k_\theta'} (x) \phi_{m'\,k_\theta'} (x') \,\phi_{n'\,k_\theta} (x'),
\end{align}
independent of $k_r'$,
\begin{equation}
 \zeta = \frac{\sqrt{2\,m}}{\ell_B\,k_r - \chi \sqrt{2\,m +\ell_B^2 k_r^2 }}\,\frac{\sqrt{2\,m'}}{\ell_B\,k_r' - \chi' \sqrt{2\,m' +\ell_B^2 {k_r'}^2 }},
\end{equation}
and $\phi_{n\,k_\theta} (x) = \psi_n (x/\ell_B + \ell_B\,k_\theta)/\sqrt{\ell_B}$.
We substitute
\begin{align}
  \frac{x}{\ell_B}  + \ell_B\,\,\frac{k_\theta + k_\theta'}{2} & \to x \nonumber \\
  \frac{x'}{\ell_B} + \ell_B\,\,\frac{k_\theta + k_\theta'}{2} & \to x'\nonumber \\
  \frac{1}{\sqrt{2}}\,\ell_B \left( k_\theta' - k_\theta \right) & \to q_\theta
\end{align}
to obtain
\begin{align}
 I_{nm}^{n'm'} =  \frac{\ell_B^2 + \xi^2}{\sqrt{\pi}\,\xi\,\ell_B} \int \di q_\theta \di x\,\di x'\, \psi_n \left( x- \tfrac{q_\theta}{\sqrt{2}} \right) \psi_m \left(x + \tfrac{q_\theta}{\sqrt{2}} \right) \nonumber \\
 \times \psi_{m'} \left(x'+ \tfrac{q_\theta}{\sqrt{2}} \right)  \psi_{n'} \left(x'-\tfrac{q_\theta}{\sqrt{2}} \right) \e^{-\frac{1}{2} \,\frac{\ell_B^2}{\xi^2} (x-x')^2 - \frac{1}{2} \xi^2 \,\delta k_\theta^2}
\end{align}
The integration over the real space coordinates is decoupled using
\begin{equation}
 \e^{-\frac{1}{2}\,\frac{\ell_B^2}{\xi^2}(x-x')^2} = \frac{1}{\sqrt{\pi}}\,\frac{\xi}{\ell_B} \int \di q_x\, \e^{-\frac{\xi^2}{\ell_B^2}\,q_x^2+\sqrt{2}\,\im\,q_x \,(x-x')},
\end{equation}
which allows to rewrite
\begin{equation}
 I_{nm}^{n'm'} = \frac{\ell_B^2 + \xi^2}{\pi\,\ell_B^2} \int \di q_x\,\di q_\theta \e^{-\frac{\xi^2}{\ell_B^2} q_x^2 - \frac{1}{2} \xi^2 \delta k_\theta^2} \mathcal{I}_{nm}^{+}\,\mathcal{I}_{n'm'}^{-}
 \label{eq:integral_nm}
\end{equation}
with the two independent integrals over real space\cite{Groenewold:1946jy}
\begin{align}
 \mathcal{I}_{nm}^{\pm}
 =	& \int \di x \,\psi_{n}  \left(x -\tfrac{q_\theta}{\sqrt{2}} \right)\,\psi_{m}  \left(x +\tfrac{q_\theta}{\sqrt{2}}, \right)\,\e^{\pm \im\,\sqrt{2}\,q_x x} \nonumber \\
 =	& \sqrt{\frac{\min (n,m)!}{\max (n,m)!}}\, \e^{-\frac{1}{2} \left( q_x^2+q_\theta^2\right)} \,L_{\min (n,m)}^{(\left| n-m\right|)}\left( q_x^2 + q_\theta^2\right) \nonumber \\
	& \times ( \pm \im\,q_x- \sgn{m-n}\,q_\theta)^{|n-m|},
\end{align}
where $L_n^{(\alpha)}$ are the Associated Laguerre polynomials.
To solve the momentum integration in Eq.~\eqref{eq:integral_nm}, we introduce polar coordinates $q_x = q\,\sin \varphi$, $q_y = q\,\cos \varphi$.
For all cases of interest that appear in Eq.~\eqref{eq:gamma_mm_full}, the product $\mathcal{I}_{nm}^{+}\,\mathcal{I}_{n'm'}^-$ depends only on the radial coordinate $q$.
Since the momenta $k_\theta, k_\theta'$ are measured from the Weyl points, the momentum difference between the states involved is $\delta k_\theta = \sqrt{2} \,q_\theta/\ell_B$ for intranode and $\delta k_\theta = \sqrt{2}\,(q_\theta \pm \tilde{b}_\theta)/\ell_B$ for internode scattering, with $\tilde{b}_\theta = \ell_B \,b_\theta/\sqrt{2}$.
Focusing on internode scattering, the integral over $\varphi$ in Eq.~\eqref{eq:integral_nm} gives
\begin{equation}
 I_{nm}^{n'm'} =  2\,(1+\gamma) \int \di q \,q\,\e^{-\gamma\,(q^2 + \tilde{b}_\theta^2 )}\,I_0 \left( 2\,\gamma\,\tilde{b}_\theta\,q \right) \,\mathcal{I}_{nm}^{+}\,\mathcal{I}_{n'm'}^-
\end{equation}
with the modified Bessel function of the first kind $I_0$ and $\gamma = \xi^2/\ell_B^2$.
We substitute $s=q^2$ and use the abbreviations $\alpha = |n-m| = |n'-m'|$ (with the restriction coming from Eq.~\eqref{eq:integral_nm}), $p=\min(n,m)$ and $l = \min(n',m')$
\begin{align}
 I_{nm}^{n'm'}
 =	& 2\,(1+\gamma)\,\sqrt{\frac{p! \, \ell!}{(p+\alpha)!\,(\ell+\alpha)!}} \,\e^{-\gamma\,\tilde{b}_\theta^2}  \\
		& \times \int \di s \,I_0 (2\, \gamma\,\tilde{b}_\theta \sqrt{s} ) \,\e^{-(1+\gamma)\,s} s^{\alpha}\,L_p^{(\alpha)} (s) \,L_l^{(\alpha)} (s) \nonumber
\end{align}
To solve the integral over $s$, we rewrite the product of Laguerre polynomials as a double sum
\begin{equation}
 L_p^{(\alpha)} (s)\,L_\ell^{(\alpha)} (s) = \sum_{i=0}^p \sum_{j=0}^\ell (-1)^{i+j} \binom{\ell+\alpha}{\ell-i} \binom{p+\alpha}{p-i} \frac{s^{i+j}}{i!\,j!}
\end{equation}
such that
\begin{align}
 & I_{nm}^{n'm'} = \frac{(1+\gamma)\,\sqrt{p! \, \ell!}\,\e^{-\gamma\,\tilde{b}_\theta^2} }{\sqrt{(p+\alpha)!\,(\ell+\alpha)!}} \sum_{i=0}^p \sum_{j=0}^\ell \frac{(-1)^{i+j}}{i!\,j!} \binom{\ell+\alpha}{\ell-i}  \nonumber \\
 &\times \binom{p+\alpha}{p-i} \int \di s \,I_0 ( 2\,\gamma\,\tilde{b}_\theta \sqrt{s} ) \,s^{\alpha+i+j}\,\e^{-(1+\gamma) s}
\end{align}
The integral over $s$ is obtained in terms of Laguerre polynomials as
\begin{align}
\int \di s &\,s^{\alpha+i+j}\,\e^{-(1+\gamma) s}\,I_0 (2\,\gamma\,\tilde{b}_\theta\,\sqrt{s} ) \nonumber \\
  &= \frac{\left( \alpha + i + j \right)!}{(1+\gamma)^{1+\alpha+i+j}}\,\e^{\frac{\gamma^2\,\tilde{b}_\theta^2}{1+\gamma}}\,L_{\alpha+i+j} \left( -\tfrac{\gamma^2\,\tilde{b}_\theta^2}{1+\gamma} \right)
\end{align}
Changing back $\tilde{b}_\theta = \ell_B \,b_\theta /\sqrt{2}$ finally gives for internode scattering
\begin{align}
 I_{nm}^{n'm'}  &= \e^{-\frac{1}{2} \frac{\xi^2 \ell_B^2}{\xi^2 + \ell_B^2}\,b_\theta^2}  \sqrt{\frac{p! \, (\ell +\alpha )!}{\ell!\,(p+\alpha)!}} \,\sum_{i=0}^p \sum_{j=0}^\ell \frac{(-1)^{i+j}}{(1+\gamma)^{\alpha+i+j}}  \nonumber \\
 \times & \binom{\ell}{j} \binom{\alpha + i+ j}{\alpha + j} \binom{p+\alpha}{p-i} \,L_{\alpha+i+j} \left( -\tfrac{1}{2} \tfrac{\xi^4}{\xi^2 + \ell_B^2}\,b_\theta^2 \right).
\end{align}
One has to be careful when numerically evaluating $I_{nm}^{n'm'}$ since the terms to be summed are exponentially large with alternating signs.
For internode scattering at $b_\theta = 0$, the integral~\eqref{eq:integral_nm} evaluates to the closed form\cite{Zwillinger:2014bd}
\begin{align}
 I_{nm}^{n'm'} 	=& \sqrt{\frac{(p+\alpha)!\,\ell!}{(\ell+\alpha)!\,p!}}\,\binom{p+\ell+\alpha}{\ell}\,\frac{\gamma^{\ell+p}}{(1+\gamma)^{\ell+p+\alpha}}   \\
		& \times \,_2 F_1 \left(-p,-\ell;-p-\ell-\alpha; \tfrac{\gamma^2-1}{\gamma^2} \right) \nonumber.
\end{align}
with the hypergeometric function~$_2 F_1$.
Ultimately, we want to compute the self-energy correction
\begin{equation}
 \Sigma^{m}_{\chi\chi'}  (\im\,\omega_n,\mathbf{k}_\parallel) = \sum_{m',k_r'} \mathcal{G}_{m' \chi'} (\im\,\omega_n , k_r') \sum_{k_\theta'} \Gamma_{mm'}^{\chi\chi'}
\end{equation}
which requires another integration over the momentum component $k_r'$.
This integral results from noting that the integrand is the imaginary part of the clean single-particle Green's function, $\mathrm{Im} G_{m\chi}^R$, which is a delta function.
The analytical results sketched here are used to compute the angle-dependence of the internode scattering time shown in the main text.

\section{Transfer matrix method for a system with Landau levels}
\label{sec:transfer_matrix}

We are looking for solutions of the Weyl equation $\mathcal{H}\,\psi = E\,\psi$, in particular at zero energy.
In a basis of left- and right-moving channels, such that $\partial \mathcal{H} /\partial p_z = v\,\sigma_z$, the Weyl equation takes the form
\begin{align}
\xi \,\partial_z\,\psi
& = \left[\im\, \frac{E - V_c - V_v\,\sigma_x\tau_x}{\hbar\omega_\xi} \sigma_z + \xi \,\mathbf{b} \cdot \left(-\sigma_y ,\sigma_x\tau_z,\im\,\tau_z \right) \nonumber \right. \\
& \left. -\frac{\xi}{\ell_B} \left(a\, \frac{\sigma_x + \im\,\sigma_y\tau_z}{\sqrt{2}} + a^\dagger \, \frac{\sigma_x- \im\,\sigma_y\tau_z}{\sqrt{2}} \right) \right] \psi,
\end{align}
with the formal solution
\begin{equation}
 \psi (z) = T(z,z') \,\psi (z')
\end{equation}
and the transfer matrix
\begin{align}
  T (z,z') =	& P_{z''} \exp \left\lbrace \int\limits_{z}^{z'} \frac{\di z''}{\xi}\,\im\,\left( \frac{E-V_c -V_v\,\sigma_x\tau_x}{\hbar\omega_\xi} \right)\sigma_z  \nonumber \right. \\
  & -\frac{\xi}{\ell_B} \left(a\, \frac{\sigma_x + \im\,\sigma_y\tau_z}{\sqrt{2}} + a^\dagger \, \frac{\sigma_x - \im\,\sigma_y\tau_z}{\sqrt{2}} \right) \nonumber \\
  & \left. + \xi \,\mathbf{b} \cdot \left(-\sigma_y ,\sigma_x\tau_z,\im\,\tau_z \right) \right\rbrace
  \label{eq:transfer_matrix}
\end{align}
where $P_{z''}$ denotes position ordering.
Two different disorder potentials are included here:
$V_c$ scatters within the same chirality and is responsible for intranode relaxation, while $V_v$ scatters between different chiralities and is responsible for internode relaxation.
The intranode potential $V_c$ does not influence the conductance, but the effect of $V_v$ is investigated in the main text of this paper.
Note that the disorder potential must be transformed to the Landau level basis,
\begin{align}
 V_{c}^{n m}		&= \int \di x\,\phi_{n\,k_y+\frac{q_y}{2}} (x)\,\phi_{m\,k_y-\frac{q_y}{2}} (x) \,V_c (x,\mathbf{q}_\parallel) \\
 V_{v\,\pm}^{n m}	&= \int \di x\,\phi_{n\,k_y+\frac{q_y}{2}} (x)\,\phi_{m\,k_y-\frac{q_y}{2}} (x) \,V_v (x,\mathbf{q}_\parallel \pm \mathbf{b}) \nonumber.
\end{align}
Instead of calculating the full transfer matrix, the system is partitioned into slices along $z$, and the transfer matrix is computed for each slice individually\cite{Bardarson:2007iu}.
These transfer matrices, connecting different sides of a finite system, are transformed to scattering matrices, connecting left- and right-moving channels\cite{Mello:1988cj,Beenakker:1997gz}.
The conductance is obtained from the concatenated scattering matrix via the Landauer formula, as elaborated in the main text.

\section{Experimental signatures of the angle-dependent magnetoresistance}
\label{sec:experimental_signatues}

Signatures of the angle-dependent magnetoresistance discussed in the main text could be observed in both Dirac and Weyl semimetals.
The ideal experiment for the identification of the strong angular dependence of $\tau_v$ on the angle between node separation and magnetic field requires a time-reversal symmetry breaking Weyl semimetal with a single pair of Weyl nodes.
In this case, the internode scattering time is directly accessible by rotating the magnetic field and measuring the conductance along the direction parallel to the magnetic field.
This may, for example, be achieved using terahertz spectroscopy that offers the advantageous possibility of contactless conductance measurements\cite{Nuss:1998hc}.

Having only two nodes is, however, not a necessary condition, and the effect is also observable in systems with more nodes.
If, for example, four Weyl nodes are arranged in a squarelike pattern in momentum space, relaxation processes between all four Weyl nodes are important.
The internode scattering time shown in Fig.~2 in the main text is, different from the case of two Weyl nodes, the same at angles $0,\pi/2,\pi$ with dips in between.
The behavior shown in Fig.~2 is restored when the two pairs of Weyl nodes get separated in momentum space, so that only scattering processes between those nodes close to each other are important.

In Dirac semimetals, a magnetic field splits a Dirac node into two Weyl nodes or nodal lines\cite{Cano:2017kc}, depending on the details of the underlying Hamiltonian.
In the case when a magnetic field splits a Dirac node into two Weyl nodes, the relevance of our work depends on the combination of the splitting direction and the relative size of the  momentum-space distance between the original Dirac nodes and their splitting. 
If the latter spacing is considerably smaller, the effect discussed in the main text will be operational.

\bibliography{correlated_disorder}

\end{document}